\documentclass[12pt, preprint]{aastex}
\begin{document}
\title{Population Boundaries for Galactic White Dwarf Binaries in LISA's
Amplitude-Frequency Domain.}
\author{Ravi Kumar Kopparapu and Joel E. Tohline}
\affil{Department of Physics and Astronomy, Louisiana State University,\\
Baton Rouge, LA - 70803.}

%%\maketitle
\begin{abstract}
Detached, inspiraling and semi-detached, mass-transferring double
white dwarf (DWD) binary systems are both expected to be important
sources for the proposed space-based gravitational-wave detector,
LISA.  The mass-radius relationship of individual white dwarf stars
in combination with the constraints imposed by Roche geometries
permit us to identify population boundaries for DWD systems in
LISA's ``absolute'' amplitude-frequency diagram.
With five key population boundaries in place, we are able to
identify four principal population sub-domains, including one
sub-domain that identifies where progenitors of Type Ia supernovae
will reside. Given one full year of uninterrupted operation, LISA
should be able to measure the rate at which the gravitational-wave
frequency $f$ and, hence, the orbital period is changing in the
highest frequency subpopulation of our Galaxy's DWD systems.  We
provide a formula by which the distance to each DWD system in this
subpopulation can be determined; in addition, we show how the masses
of the individual white dwarf stars in mass-transferring systems may
be calculated.
%White dwarf  binary systems (inspiralling and mass transferring
%binaries) are  an important source for the proposed space based
%gravitational wave detector LISA (Laser Interferometer Space
%Antenna). Theoretical constraints on the properties of the white
%dwarf stars along with the operational time of LISA (assumed one
%year here) may allow us to confine their population in the LISA
%amplitude-frequency space.
%%and a measurement of the rate
%%of change of amplitude ($h^{\prime}$) and frequency ($f^{\prime}$)
%%can give the individual binary component masses.
%In this paper, we discuss in detail these limits and apply them to
%white dwarf binary systems within our galaxy ($\approx$ 10Kpc).
%Measurement of the rate at which the frequency and amplitude changes
%will allow us to determine the binary system parameters and also
%will indicate the physical mechanism governing the evolution.
%Bounds on the exitence of various interesting astrophysical sources
%such as Type Ia supernovae, AM CVn systems in the
%amplitude-frequency domain of LISA is also considered.
\end{abstract}

\section{Introduction}
Double white dwarf (DWD) binaries are considered to be one of the
most promising sources of gravitational waves for
LISA,\footnote{http://lisa.nasa.gov} the proposed Laser
Interferometer Space Antenna \citep{FB84,EIS87,Bender98}. If, as has
been predicted \citep{IT84, IT86}, close DWD pairs are the end
product of the thermonuclear evolution of a sizeable fraction of all
binary systems, then DWD binaries must be quite common in our Galaxy
and the gravitational waves (GW) emitted from these systems may be a
dominant source of background noise for LISA in its lower frequency
band,  $f \lesssim 3 \times 10^{-3} ~\mathrm{Hz}$ \citep{Hils90,
Corn03}. DWD binaries are also believed to be (one of the likely)
progenitors of Type Ia supernovae \citep{IT84,B95,T05} in situations
where the accreting white dwarf exceeds the Chandrasekhar mass
limit, collapses toward nuclear densities, then explodes.  Because
its instruments will have sufficient sensitivity to detect GW
radiation from close DWD binaries throughout the volume of our
Galaxy, LISA will provide us with an unprecedented opportunity to
study this important tracer of stellar populations and it will
provide us with a much better understanding of the formation and
evolution of close binary systems in general. Clearly, a
considerable amount of astrophysical insight will be gained from
studying the DWD population as a guaranteed source for LISA.

Broadly speaking, our Galaxy's DWD binary population should be
dominated by systems that are in two distinctly different
evolutionary phases: An ``inspiral'' phase, during which both stars
are detached from their respective Roche lobes; and a semi-detached,
``stable mass-transfer'' phase during which the less massive star
fills its Roche lobe and is slowly transferring mass to its more
massive companion.  While DWD binaries may encounter other
interesting evolutionary phases -- for example, a phase of so-called
common envelope evolution, or a phase of rapid, unstable mass
transfer -- the inspiral and stable mass-transfer phases are
expected to dominate the population because they are especially
long-lived.  It should be noted that we already have a handle on the
size of the galactic population of DWD binaries from optical, UV,
and x-ray observations.  In the immediate solar
neighborhood, there are 18 systems\footnote{Three models \citep{Cropper1998, 
Wu2002, MS2002} have been proposed to 
determine the nature of two controversial candidate systems
 (RX J0806+15 and V407 Vul) out of these 18, which can change the number of 
known AM CVn systems between 16 and 18.}
 \citep{Nel05, AND05, ROE05} known to be  
 undergoing a phase of stable mass transfer (AM CVn being the prototype) and the
 ESO SN Ia Progenitor SurveY (SPY) has detected nearly 100 detached DWD systems
\citep{Nap04}. At present, orbital periods and the component masses for 24
 detached DWD systems have been determined
 (see Table 3 of \cite{Nelet05} and references therein), five of which come
from the SPY survey.
%  Within this
%sample, systems undergoing a phase of stable mass transfer -- AM CVn
%being the prototype -- dominate over detached systems, presumably
%because the photon flux generated by associated accretion processes
%makes the mass-transferring systems more readily detectable.

During both of these relatively long-lived evolutionary phases, a
DWD system's orbital period (and corresponding GW frequency) changes
on a time scale that is governed by the rate at which angular
momentum is being lost from the system due to gravitational
radiation, that is, the so-called ``chirp'' time scale (see the
discussion associated with Eq. \ref{tau_chirp}, below).  While the
system is detached, the orbital separation slowly decreases so the
system should emit a GW signal with a characteristic ``chirp''
signature, that is, the frequency and amplitude of the GW signal
should monotonically increase with time. During a phase of stable
mass transfer, however, the orbital separation steadily increases so
the GW signal should exhibit an inverse-chirp character where by
its frequency and amplitude should steadily decrease with time.  The
primary objective of our present study is to analyze the imprint
that both of these relatively long-lived phases of evolution will
have on our Galaxy's DWD binary population, as viewed by LISA.

LISA's capabilities as a GW detector are usually discussed in the
context of the $\log(h)-\log(f)$ diagram, where the GW frequency $f$
is measured in Hertz, and the GW amplitude $h$ is a dimensionless
``strain'' (generally quoted per $\sqrt{\mathrm{Hz}}$, reflecting
the frequency resolution of the data stream).  A useful analogy can
be drawn between this amplitude-frequency diagram and the astronomy
community's familiar color-magnitude (CM) diagram. Directly from
photometric measurements, astronomers can produce a CM diagram that
is based on the apparent brightness (the apparent magnitude $m$) of
various sources.  However, a determination of the intrinsic
brightness of each source must await the determination of the distance
$r$ to each source and the corresponding conversion of each measured
value of $m$ to an absolute magnitude $M$.  LISA's measurement of
$\log(h)$ for a given astrophysical source is analogous to a
measurement of $m$; it only tells us how bright the GW source
appears to be on the sky.  A determination of the intrinsic
brightness of each GW source must await a determination of the
source distance $r$ and the corresponding conversion of each
measured value of (the apparent brightness) $h$ to a quantity that
represents the ``absolute'' brightness of the source.  For LISA
sources, the relevant quantity (analogous to $M$) is $\log(rh)$.
Astronomers realize that the underlying physical properties of
stars, their evolution, and their relationship to one another in the
context of stellar populations can only be ascertained from a CM
diagram if $M$, rather than $m$, is used to quantify stellar
magnitudes. By analogy, it should be clear that the underlying
physical properties of DWD systems, their evolution, and their
relationship to one another in the context of stellar populations
can be ascertained only if the observational properties of such
systems are displayed in a $\log(rh) - \log(f)$ diagram, rather than
in a plot of $\log(h)$ versus $\log(f)$.  For this reason, our
discussion of DWD systems will be presented in the context of this
more fundamental, but rather under-utilized, ``absolute''
amplitude-frequency domain.

As we investigate the evolution of DWD systems across LISA's
``absolute'' amplitude-frequency domain, we will utilize a
simplified description of the two long-lived evolutionary phases
mentioned above.  We will assume (1) all orbits are circular; (2)
the orbital frequency is related to the orbital separation via
Kepler's third law; (3) the spin of both stars can be ignored so
that each system's total angular momentum is given by the point-mass
expression for orbital angular momentum; (4) the total mass of each
system is conserved; and (5) angular momentum is lost from each
system only via the radiation of gravitational waves and that the
rate of angular momentum loss is correctly described by a 
quadrupole radiation formula.
%low-order,
%post-Newtonian analysis of general relativity. 
 Simplification \#4
means, for example, that after the low-mass white dwarf comes into
contact with its Roche lobe, we will assume that each DWD
system evolves along a ``conservative'' mass transfer (CMT)
trajectory, and simplification \#5 means that we will be ignoring
effects that might arise due to direct-impact accretion
\citep{MNS,GPF06}. A more thorough analysis that removes some or all
of these simplifications is likely to provide additional valuable
insight into the evolution of DWD populations; \cite{SVN05}, for
example, have expressed concern that the time-rate-of-change of the
GW frequency for many of LISA's most interesting sources will not be
correctly interpreted without a proper treatment of tides.  In this
context, our analysis should be viewed as an important first step in
what is likely to be a long-term study of the evolution of the DWD
binary population across LISA's ``absolute'' amplitude-frequency
domain.  At the outset, we acknowledge that our understanding of
this subject has benefitted significantly from the insight of others
whose work has preceded ours; most notably, we recognize the
insightful publications by \cite{Paczynski67}, \cite{Faulkner71},
\cite{EIS87}, \cite{WI87}, \cite{MNS}, and \cite{GPF06}.

\section{Parameterization}
%A detached double white dwarf (DWD) binary can evolve by loosing orbital
%angular momentum due to gravitational radiation and the gravitational-wave
%amplitude produced as result can be approximated with
In the quadrupole approximation \citep{PM63,Thorne87,FC93}, the
time-dependent gravitational-wave strain, $h(t)$, generated by a
point mass binary system in circular orbit has two polarization
states.  The plus and cross polarizations of $h(t)$ generically take
the respective forms,\footnote{Throughout this paper when we refer
to experimental measurements of $h$, we will assume that the binary
system is being viewed ``face on'' so that the measured peak-to-peak
amplitude of the two polarization states are equal and at their
maximum value, given by $h_\mathrm{norm}$. If the orbit is inclined
to our line of sight, the inclination angle can be determined as
long as a measurement is obtained of both polarization states as
shown, for example, by \cite{FC93}.  Because our discussion focuses
on Galactic DWD binaries, we ignore the effects of cosmological
expansion.}
%will also assume that the effects of
%cosmological expansion on measured signal strengths is negligible.}
%\begin{equation}\label{generic_h}
$h_{+} = h_\mathrm{norm} \cos[\phi(t)]$ and
% \, \qquad \mathrm{and} \qquad
$h_{\times} = h_\mathrm{norm} \sin[\phi(t)]$, 
%\, ,
%\end{equation}
where the time-dependent phase angle,
\begin{equation}
\phi(t) = \phi_0+ 2\pi\int f(t)dt \, , \label{phaseDefinition}
\end{equation}
where $\phi_0$ is the phase at time $t=0$,
$f=\Omega_\mathrm{orb}/\pi$ is the frequency of the gravitational
wave measured in Hz, $\Omega_\mathrm{orb}$ is the angular velocity
of the binary orbit given in radians per second, and the
characteristic amplitude of the wave,
\begin{equation}\label{hnorm}
h_\mathrm{norm} = \frac{G}{rc^4} ~\frac{4 \Omega_\mathrm{orb}^{2}
M_{1}M_{2} a^{2}}{(M_1 + M_2)} \, ,
\end{equation}
where $G$ is the gravitational constant, $c$ is the speed of light,
$r$ is the distance to the source, $M_1$ and $M_2$ are the masses of
the two stars, and $a$ is the distance between the stars. If the
principal parameters of the binary system do not change with time,
then $f$ and $h_\mathrm{norm}$ will both be constants and the phase
angle $\phi$ will vary only linearly in time, so the source will
emit (monochromatic) ``continuous-wave'' radiation. If, however, any of
 the binary
parameters --- $M_1$, $M_2$, $a$, or $\Omega_\mathrm{orb}$
--- vary with time, then $h_\mathrm{norm}$ and/or $f$ will also vary
with time in accordance with the physical process that causes the
variation.  Here we will only be considering the evolution of DWD
systems in which the basic system parameters vary on a timescale
that is long compared to $1/f$.

We know from the mass-radius relationship for white dwarfs (see the
discussion associated with Eq.~\ref{mass_radius}, below) that the
less massive star in a DWD binary will always have the larger
radius.  Therefore, in a DWD system that is undergoing mass
transfer, we can be certain that the less massive star is the
component that is filling its Roche lobe and is transferring
(donating) mass to its companion (the more massive, accretor). With
this in mind, throughout the remainder of our discussion we will
identify the two stars by the subscripts $d$ (for donor) and $a$
(for accretor), rather than by the less descript subscripts $1$ and
$2$, and will always recognize that the subscript $d$ identifies the
less massive star in the DWD system.  This notation will be used
even during evolutionary phases (such as a GR-driven inspiral phase)
when the two stars are detached and therefore no mass-transfer is
taking place. Furthermore, we will frequently refer to the total
mass of the system,
%\begin{equation}
$M_\mathrm{tot} \equiv M_d + M_a$ % \, ,
%\end{equation}
and the mass ratio,
\begin{equation}\label{q_def}
q \equiv \frac{M_d}{M_a} \, ,
\end{equation}
which will necessarily be confined to the range $0 < q \leq 1$
because $M_d \leq M_a$. Also, it will be understood that the
limiting mass for either white dwarf is the Chandrasekhar mass,
$M_\mathrm{ch} = 1.44 M_\mathrm{\sun}$.

As mentioned above, throughout this investigation we will assume
that Kepler's third law provides a fundamental relationship between
the angular velocity and the separation of DWD binaries, that is,
\begin{equation}\label{Kepler}
\Omega_\mathrm{orb}^2 = \frac{G M_\mathrm{tot}}{a^3} \, .
\end{equation}
%%(See \cite{Motl01} for the calculation of a correction to this
%%relation that may be incorporated to improve the treatment of
%%tidally distorted, mass-transferring systems.)
Relation (\ref{Kepler}) allows us to replace either
$\Omega_\mathrm{orb}$ or $a$ in favor of the other parameter in
Eq.~(\ref{hnorm}). Furthermore, we will find it useful to
interchange one or both of these parameters with the binary system's
orbital angular momentum 
%which, via the above relations,
%can be expressed in any of the following forms:
\begin{equation}\label{JorbDefined}
J_\mathrm{orb} \equiv \frac{M_aM_d a^2\Omega_\mathrm{orb}}{M_\mathrm{tot}} = \biggl(\frac{G^2
M_\mathrm{tot}^5}{\pi f}\biggr)^{1/3} Q \, ,
\end{equation}
where,
\begin{equation}
Q \equiv \frac{q}{(1+q)^2} \, , \label{Q_defined}
\end{equation}
is the ratio of the system's reduced mass to its total mass.
For our future discussion, it is useful to express the gravitational wave 
amplitude $h_\mathrm{norm}$ and frequency $f$ in terms of $J_\mathrm{orb}$
 and $Q$. So,
\begin{eqnarray}
h_\mathrm{norm} &=& \frac{4G^3}{rc^4} \frac{M_\mathrm{tot}^5 Q^3}{J_\mathrm{orb}^{2}} \, ,
\label{hnorm_J_Q} \\
f &=& \frac{G^2}{\pi} \frac{M_\mathrm{tot}^5 Q^3}{J_\mathrm{orb}^{3}} .
\label{f_J_Q}
\end{eqnarray}
% Table \ref{TemplateTable} summarizes how the frequency $f$ and
%``absolute'' amplitude $rh_\mathrm{norm}$ of the gravitational-wave
%strain can be expressed in terms of $M_\mathrm{tot}$, $Q$, and
%either $J_\mathrm{orb}$, $a$, or $\Omega_\mathrm{orb}$.
  We note as
well that the so-called ``chirp mass'' $\mathcal{M}$ of a given
system \citep{FC93} is obtained from $M_\mathrm{tot}$ and $Q$ via
the relation,
\begin{equation}
\mathcal{M} = M_\mathrm{tot}Q^{3/5} \, .
\end{equation}
%\clearpage
%\begin{deluxetable}{cccc}
%\tablecaption{Template Formulae} \tablewidth{0pt}
%
%\tablehead{ \colhead{Specify:} & \colhead{$J_\mathrm{orb}$} &
%\colhead{$a$ }  & \colhead{$\Omega_\mathrm{orb}$}
%\\
%\colhead{(1)} & \colhead{(2)} & \colhead{(3)} & \colhead{(4)}}
%
%\startdata $rh_\mathrm{norm}$
%       & $\frac{4}{c^4}
%         G^3 M_\mathrm{tot}^5 J_\mathrm{orb}^{-2} Q^3$
%     & $\frac{4}{c^4}G^2 M_\mathrm{tot}^2
%     a^{-1} Q$
%     & $\frac{4}{c^4}(GM_\mathrm{tot})^{5/3}
%         \Omega_\mathrm{orb}^{2/3} Q$   \\
%~& ~ & ~ & ~ \\
%$f$   & $\frac{1}{\pi} G^2 M_\mathrm{tot}^5 J_\mathrm{orb}^{-3}
%Q^3$   & $\frac{1}{\pi} (GM_\mathrm{tot})^{1/2} a^{-3/2}$  & $\frac{1}{\pi} \Omega_\mathrm{orb}$   \\
%\enddata\label{TemplateTable}
%\end{deluxetable}
%\clearpage
\section{Evolution of DWD Binaries in the Amplitude-Frequency Domain}
{\label{theory_section.}}

\subsection{Trajectories and Termination Boundaries}
%\subsubsection{Inspiral due to Gravitational Radiation}

\begin{figure}[!hbp|t]
\vspace*{-25mm}
\centering
%\epsscale{0.4}
\plotone{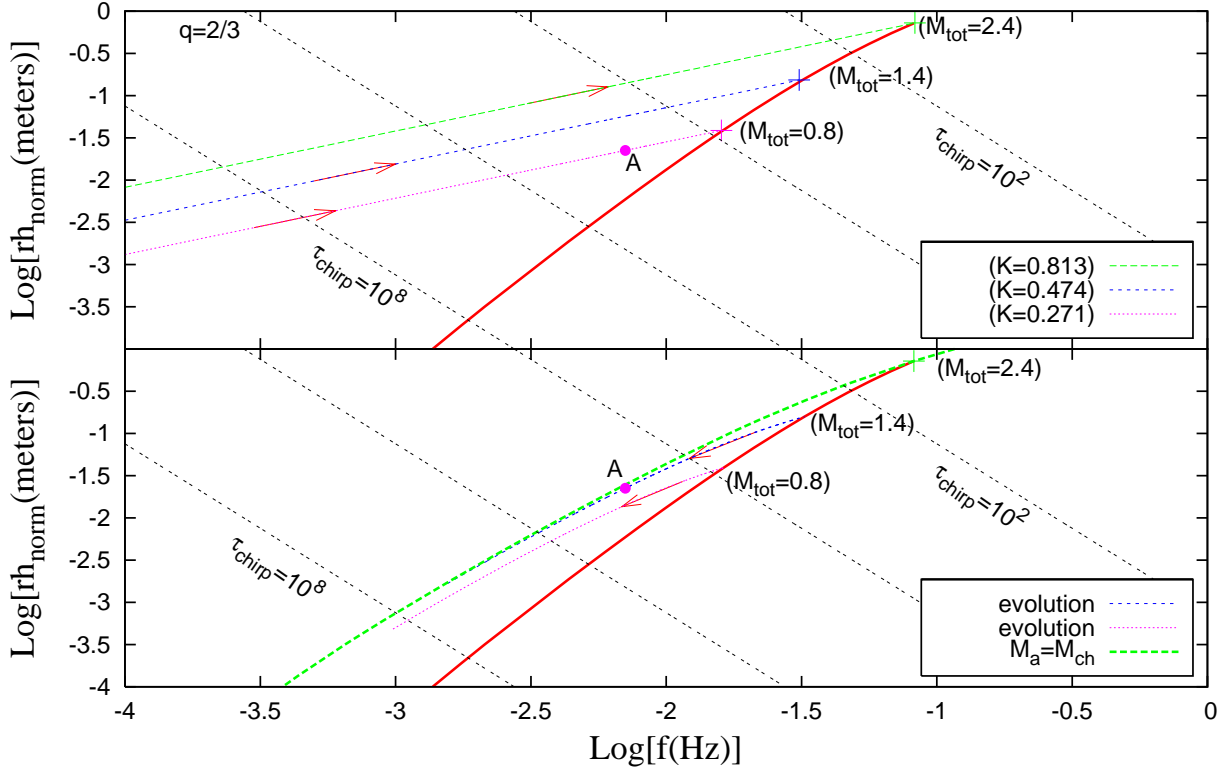}
\hfill \caption{{\it Top panel:} Example evolutionary trajectories
(lines with arrows pointing to the upper-right) for detached, DWD
systems that are undergoing a GR-driven inspiral are displayed in
this $\log(rh_\mathrm{norm}) - \log f$ diagram, where
$rh_\mathrm{norm}$ is specified in meters and $f$ is specified in
Hz. The three trajectories represent systems having dimensionless
mass parameters $K = 0.813$, $0.474$, and $0.271$ as indicated;
assuming a mass ratio $q=2/3$ for all three systems, this
corresponds to total system masses of $2.4, 1.4$, and $0.8 M_\odot$,
respectively. The termination point of each detached, inspiral
trajectory is marked by a plus symbol.  The solid red curve passing
through the plus symbols forms a termination boundary for
inspiraling systems that have a mass ratio $q = 2/3$ but a wide
range of total masses, $0.06 M_\odot \leq M_\mathrm{tot} \leq 2.4
M_\odot$; for this mass ratio, $2.4 M_\odot$ represents the maximum
allowed total system mass because $M_a = M_\mathrm{ch}$. Dotted
black lines having a slope of $-2$ depict various ``chirp
isochrones;'' as indicated, the isochones identify systems whose
characteristic evolutionary time scales, $\tau_\mathrm{chirp}$,
range from $10^2 - 10^8 \textrm{yrs}.$ {\it Bottom panel:} In the
same $\log(rh_\mathrm{norm}) - \log f$ parameter space, evolutionary
trajectories (curves with arrows pointing to the lower-left) are
displayed for DWD systems that are undergoing a stable phase of CMT.
The two illustrated trajectories have been drawn for systems that
begin their mass-transfer evolutions with a system mass ratio $q_0 =
2/3$ -- hence the trajectories begin at the termination points of
the respective inspiral trajectories shown above -- but they have
different total system masses: $M_\mathrm{tot} = 1.4 M_\odot$ (blue
dotted curve) and $M_\mathrm{tot} = 0.8 M_\odot$ (pink dotted
curve).  The (green) thick dashed curve represents a low-frequency
boundary for semi-detached, mass-transferring DWD binaries because,
in order for a system to move beyond this boundary, the mass of the
accretor $M_a$ would have to exceed $M_\mathrm{ch}$.  The single
point marked ``A'' in both panels $[~\log(rh_\mathrm{norm}), \log f
~] = [ -1.65, -2.15 ]$ can represent DWD systems having multiple
$(q,M_\mathrm{tot})$ pairs. } {\label{logh_logf_q0.66}}
\end{figure}

Detached DWD binaries slowly inspiral toward one another as they
lose orbital angular momentum due to gravitational radiation.
Assuming that $M_\mathrm{tot}$ and $q$ remain
constant during this phase of evolution, Eqs.(\ref{hnorm_J_Q}) and 
(\ref{f_J_Q}) can be combined to give,
% the expressions given
%in column 2 of Table \ref{TemplateTable} show that both the
%frequency and amplitude of the emitted GW signal will increase as
%the system's orbital angular momentum decreases. Combining these
%expressions in a way that cancels out the dependence on
%$J_\mathrm{orb}$, we obtain,
\begin{eqnarray}
rh_\mathrm{norm} &=& \biggl[ \frac{2^5\pi^2}{c^2} \biggl(
\frac{GM_\mathrm{ch}}{c^2} \biggr)^5  K^5 f^2 \biggr]^{1/3} = 5.38
~[K^5 f^2]^{1/3}~\mathrm{meters} \, , \label{h_f_relationship}
\end{eqnarray}
where
%%$M_\mathrm{ch} = 1.44~M_\odot$ is the Chandrasekhar mass and
the dimensionless mass parameter,
\begin{eqnarray}\label{K_definition}
K \equiv 2^{1/5}\biggl( \frac{\mathcal{M}}{M_\mathrm{ch}} \biggr) =
2^{1/5} \biggl( \frac{M_\mathrm{tot}}{M_\mathrm{ch}}  \biggr)
Q^{3/5} = \biggl(\frac{M_\mathrm{a}}{M_\mathrm{ch}}\biggr) \biggl(
\frac{2 q^3}{1+q}\biggr)^{1/5} \, ,
\end{eqnarray}
has been defined such that it acquires a maximum value of unity in
the limiting case where $M_d = M_a = M_\mathrm{ch}$; otherwise, $0 <
K < 1$. (We note that in the limiting case of $K=1$, the chirp mass
of the system is $\mathcal{M} = 0.871 M_\mathrm{ch} = 1.25
M_\odot$.) From expression (\ref{h_f_relationship}), we see that the
trajectory of an inspiraling, detached DWD binary in the
amplitude-frequency diagram can be determined without specifying
precisely the rate at which angular momentum is lost from the
system. Specifically, because $d\ln(rh_\mathrm{norm})/d\ln f = 2/3$,
trajectories of inspiraling DWD binaries will be straight lines with
slope $2/3$ in a plot of $\log(rh_\mathrm{norm})$ versus $\log f$.
By way of illustration, inspiral trajectories for systems having
three different total masses ($2.4~M_\odot$, $1.4~M_\odot$, and
$0.8~M_\odot$) but all having the same $q=2/3$ mass ratio are
displayed in the top panel of Figure \ref{logh_logf_q0.66}. 
%As the
%arrows indicate, during the GR-driven inspiral, evolution is up and
%to the right in this  diagram.

%\subsubsection{Roche-Lobe Contact}\label{Sec:RLcontact}
%\subsubsection{Evolution to lower frequencies due to conservative 
%mass transfer}\label{Sec:RLcontact}

The detached inspiral phase of the evolution of a DWD binary will
terminate when the binary separation $a$ first becomes small enough
that the less massive white dwarf fills its Roche lobe.  From 
Eggleton's mass-radius relationship for zero-temperature white dwarfs, as 
quoted by \cite{VR88} and also in \cite{MNS}, 
 we know that the radius of the donor $R_d$ is,
\begin{eqnarray}
\frac{R_d}{R_\mathrm{\sun}} = 0.0114
\biggl[\biggl(\frac{M_d}{M_\mathrm{ch}} \biggr)^{-2/3} -
\biggl(\frac{M_d}{M_\mathrm{ch}}\biggr)^{2/3}\biggr]^{1/2}
\biggl[1+3.5\biggl(\frac{M_d}{M_\mathrm{p}}\biggr)^{-2/3}+
\biggl(\frac{M_d}{M_\mathrm{p}}\biggr)^{-1}\biggr]^{-2/3} \, ,
\label{mass_radius}
\end{eqnarray}
where $M_p \equiv 0.00057~M_\odot$.  Furthermore, from \cite{EGG} we
find that the Roche-lobe radius $R_L$ is,
\begin{eqnarray}
R_\mathrm{L} \approx a \biggl[ \frac{0.49 \ q^{2/3}}{0.6 \ q^{2/3} +
\ln(1 + q^{1/3})}\biggr]. 
%=\frac{J_\mathrm{orb}^2}{GM_\mathrm{tot}^3}\frac{(1+q)^4}{q^2}
%\biggl[ \frac{0.49 \ q^{2/3}}{0.6 \ q^{2/3} + \ln(1 +
%q^{1/3})}\biggr] \, . 
\label{roche_radius}
\end{eqnarray}
The orbital separation -- and the corresponding GW amplitude
$rh_\mathrm{norm}$ and frequency $f$ -- at which the inspiral phase
terminates can therefore be determined uniquely for a given donor
mass $M_d$ and system mass ratio $q$ by setting $R_d = R_L$ and
combining expressions (\ref{mass_radius}) and (\ref{roche_radius})
accordingly.  The termination points of the three inspiral
trajectories --- marked by plus symbols in the top panel of Figure
\ref{logh_logf_q0.66} ---  have been calculated in this manner. The
curve connecting the sequence of plus symbols in Figure
\ref{logh_logf_q0.66} traces out the locus of points that define the
termination points of the detached inspiral phase of numerous other
DWD systems that have mass ratios $q=2/3$ but that have values of
$M_\mathrm{tot}$ ranging from $2.4~M_\odot$ to $0.06~M_\odot$.

 As a DWD system fills its Roche lobe and starts transferring mass to its
companion, it evolves to lower frequencies and amplitudes. Without knowing the
 precise rate at which this phase of mass transfer proceeds, we can
map out the evolutionary trajectory of various sytems in the
$\log(rh_\mathrm{norm}) - \log f$ diagram if we assume that the
system's total mass is conserved and the donor's radius is marginally in 
contact with its Roche lobe. By way of illustration, the bottom panel of
Figure \ref{logh_logf_q0.66} shows two stable, conservative
mass-transfer (CMT) trajectories: The (blue) dashed trajectory is for a
 system of mass
$M_\mathrm{tot} = 1.4~M_\odot$; the (pink) dotted trajectory is for
a system of mass $M_\mathrm{tot} = 0.8~M_\odot$.
% As the arrows
%indicate, along both mass-transfer trajectories evolution is down
%and to the left in this amplitude-frequency diagram.
 We have assumed
that both of these systems began the mass-transfer phase of their
evolution with an initial mass ratio $q_0 = 2/3$. Hence, the
starting point of both trajectories lies on the termination boundary
for inspiralling systems having mass ratios of $q = 2/3$. For systems with
$M_\mathrm{tot} > M_\mathrm{ch}$, the mass of the
accretor will exceed $M_\mathrm{ch}$ when $q$ drops
below the value,
\begin{equation}
q_\mathrm{ch}\equiv \frac{M_\mathrm{tot}}{M_\mathrm{ch}} - 1 \,,
\qquad \mathrm{for}~M_\mathrm{tot} > M_\mathrm{ch} \, .
\label{qChandra}
\end{equation}
With the expectation that something catastrophic ({\it e.g.}, a Type
Ia supernova explosion) will occur when this happens, 
 it is reasonable to assume that CMT
trajectories with $M_\mathrm{tot}>M_\mathrm{ch}$ will terminate at a
point in the amplitude-frequency diagram that is marked by $q_\mathrm{ch}$.
The locus of points that
is defined by the termination points of these trajectories defines
another interesting astrophysical boundary in LISA's ``absolute''
amplitude-frequency diagram.  This termination boundary has been
drawn as a thick, (green) dashed curve in the bottom panel of Figure
\ref{logh_logf_q0.66}.
%\subsubsection{Boundaries in the Amplitude-Frequency Domain}

The inspiral trajectory drawn for $K=0.813$ ($M_\mathrm{tot} =
2.4~M_\odot$) and the curve marking the termination of various
inspiral trajectories in the top panel of Figure
\ref{logh_logf_q0.66} define boundaries in the amplitude-frequency
domain outside of which no DWD system should exist if it has a mass
ratio $q \leq 2/3$.
%  As explained above, DWD evolutionary
%trajectories are expected to ``bounce'' off of the high-frequency
%``termination'' boundary and thereafter move toward lower
%frequencies because, at that boundary, mass transfer begins.  And to
%exist above the $K = 0.813$ inspiral trajectory, the more massive
%star would have to have a mass $M_a > M_\mathrm{ch}$ if $q = 2/3$.
Analogous domain boundaries can be constructed readily for other
values of $q$. 
%(see Figure
%\ref{curved_parameter_space} for examples).
 For each value of $q$,
the shapes of the bounding curves are roughly the same as shown in
the top panel of Figure \ref{logh_logf_q0.66}, but for higher (lower) values
of $q$ the right-hand termination boundary shifts to higher (lower)
frequencies and the limiting inspiral trajectory (set by the parameter $K$)
% (set by a higher
%value of the mass parameter $K$)
 shifts to higher (lower) strain amplitudes.
%for lower values of $q$ the termination boundary shifts to lower
%frequencies and the limiting inspiral trajectory shifts to lower
%strain amplitudes.
  Given our present understanding of the structure
of white dwarfs, it seems extremely unlikely that any DWD binary
systems can exist outside of the domain that is defined by the
bounding curves for systems with $q=1$ (see, for example, 
 Figure \ref{conclusion_figure}).
%{\bf[When Ravi extends this curve to touch the lower horizontal
%axis, this minimum mass value needs to be recalculated
%accordingly.]}
\clearpage
\thispagestyle{empty}
\clearpage

\subsection{Time-Dependence}

Up to this point, we have described key features of DWD evolutionary
trajectories in LISA's ``absolute'' amplitude-frequency diagram
without referring to the rate at which the evolution of any given
system proceeds. Here we investigate the time scales on which
significant changes in various system parameters and, as a
consequence, the rate at which measurable changes in the GW
signature occur.
% Drawing on the expressions given in column 2 of
%Table \ref{TemplateTable},
Using Eqs.(\ref{hnorm_J_Q}) and (\ref{f_J_Q}), and assuming $M_\mathrm{tot}$ 
as constant, 
 we can write the time-rate-of-change of
the amplitude and frequency, for both the inspiral and CMT phase, as
follows:
%\begin{eqnarray}
%\frac{d\ln h_\mathrm{norm}}{dt} &=& 5 \frac{\partial\ln
%M_\mathrm{tot}}{\partial t} - 2 \frac{\partial\ln
%J_\mathrm{orb}}{\partial t}
%+ 3 \frac{\partial\ln Q}{\partial t} \, ; \\
%\frac{d\ln f}{dt} &=& 5 \frac{\partial\ln M_\mathrm{tot}}{\partial
%t} - 3 \frac{\partial\ln J_\mathrm{orb}}{\partial t} + 3
%\frac{\partial\ln Q}{\partial t} \, .
%\end{eqnarray}
%Assuming that the binary system's total mass is conserved during
%either the GR-driven inspiral phase or a phase of stable CMT, we can
%drop the first term on the right-hand-side of both of these
%equations to obtain,
\begin{eqnarray}
\frac{d\ln h_\mathrm{norm}}{dt} = - 2 \frac{\partial\ln
J_\mathrm{orb}}{\partial t} + 3 \frac{\partial\ln Q}{\partial t} \,
; \qquad \frac{d\ln f}{dt} = - 3 \frac{\partial\ln
J_\mathrm{orb}}{\partial t} + 3 \frac{\partial\ln Q}{\partial t} \,.
\label{dlnf_MT}
\end{eqnarray}
%These expressions can be used to deduce the rate of change of
%$h_\mathrm{norm}$ and $f$ during a phase of CMT when the system mass
%ratio (and, hence, the function $Q$) is steadily changing and,
%simultaneously, the system is losing angular momentum due to the
%radiation of gravitational waves. On the other hand, during a phase
%of GR-driven inspiral, both stars in the DWD binary are detached
%from their respective Roche lobes so, although orbital angular
%momentum is being steadily lost from the system, $q$ (hence, $Q$) is
%not changing and the following, even simpler expressions apply:
%\begin{eqnarray}
%\frac{d\ln h_\mathrm{norm}}{dt} \approx - 2 \frac{\partial\ln
%J_\mathrm{orb}}{\partial t} \, ; \qquad \frac{d\ln f}{dt} \approx -
%3 \frac{\partial\ln J_\mathrm{orb}}{\partial t}  \,.
%\label{dlnf_inpiral}
%\end{eqnarray}

%%\subsection{Determination of $q$ and $M_\mathrm{tot}$}
%\subsubsection{GR-Driven Inspiral}

During the inspiral phase of DWD binary evolution
 $\partial\ln Q/\partial t = 0$, so 
the evolution is driven entirely by the loss of angular momentum due to
gravitational radiation.
 According to \cite{PM63} (see also
\citet{MTW}), the time-dependent behavior of $J_\mathrm{orb}$ is
described by the relation,
% starting at time $t=0$ from any orbital separation
%$a_0$ -- and corresponding orbital angular momentum $J_0$, strain
%amplitude $h_0$, and GW frequency $f_0$ -- to a high degree of
%precision the time-dependent behavior of $J_\mathrm{orb}$ is
%described by the relation,
\begin{equation}\label{JofTviaGR}
J_\mathrm{orb}(t) = J_0 (1-t/\tau_\mathrm{chirp})^{1/8} \, ,
\end{equation}
where the inspiral evolutionary time scale is,
\begin{eqnarray}
\tau_\mathrm{chirp} &\equiv& \frac{5}{256} \frac{c^5 a^4}{G^3
M_\mathrm{tot}^{3}} \biggl[\frac{(1+q)^2}{q}\biggr] = \frac{ 5}{64
\pi^2 } \biggl( \frac{c}{rh f^2} \biggr) \, . \label{tau_chirp}
\end{eqnarray}
Hence, 
\begin{eqnarray}
\frac{\partial \ln J_\mathrm{orb}}{\partial t} \approx -\frac{1}{8 \tau_\mathrm{chirp}}.
\label{dJorb_dt}
\end{eqnarray}

For the CMT phase, $\partial \ln Q/\partial t \ne 0$.
Hence both the terms on the right-hand side of Eq. (\ref{dlnf_MT}) 
affect the evolution.
 From the work of \cite{WI87} and \cite{MNS}, we deduce that during 
 a phase of stable CMT (see \cite{K2006} for a detailed derivation),
\begin{equation}
\frac{\partial \ln Q}{\partial t} \approx -\biggl(\frac{1-q}{4 \Delta \zeta}\biggr)\frac{1}
{\tau_\mathrm{chirp}} \, ,
 \label{tau_mt}
\end{equation}
where $\Delta\zeta(M_\mathrm{tot}, q) \equiv \zeta_\mathrm{d} - 
\zeta_\mathrm{R_\mathrm{L}}$
is a parameter that is of order unity. The quantities $\zeta_\mathrm{d}$
and $\zeta_\mathrm{R_\mathrm{L}}$ represent the change in the donor's
 radius and Roche
lobe radius, respectively,  as a function of its mass \citep{MNS}\footnote{
Here we use the notation $\zeta_\mathrm{d}$ to represent the change in the
radius of the donor with respect to its mass, whereas \cite{MNS} use
$\zeta_\mathrm{2}$ indicating the donor as the secondary star. Also,
 they use a slightly different definition for
$\zeta_\mathrm{R_\mathrm{L}}$.
The relation between their's ($\zeta_\mathrm{r_\mathrm{L}}$) and
 our's ($\zeta_\mathrm{R_\mathrm{L}}$) is : $\zeta_\mathrm{r_\mathrm{L}} =
\zeta_\mathrm{R_\mathrm{L}} + 2 (1 - q)$. }.
It should be emphasized that 
%a phase of stable CMT can occur only if $\Delta\zeta$ is positive and
 the timescale on which DWD systems evolve during both the inspiral and CMT 
phases is
$\sim  \tau_\mathrm{chirp}$, as indicated by Eqs.~(\ref{dJorb_dt}) and 
(\ref{tau_mt}).
It is for this reason that we have drawn
various ``chirp isochrones'' in both
panels of Figure \ref{logh_logf_q0.66}; as can be ascertained from
Eq.(\ref{tau_chirp}), each isochrone depends only on the product of
$rh$ and $f^2$ and, hence, has a slope of $-2$ in the figure panels.
%Table \ref{Table:DeltaZeta} below lists the values of
% $q_\mathrm{crit}$ that
%correspond to five separate values of $M_\mathrm{tot}$.

In practice, for a given source, LISA will be unable to measure changes
 in the strain amplitude $h$ at the levels predicted by
expression (\ref{dlnf_MT}) because variations in $h$ do not accumulate
secularly over time and, in particular, will not contribute to
the observed phase of the signal. 
 Since, as indicated in Eq.(\ref{phaseDefinition}), the phase
depends on the time-rate-of-change
 of the GW frequency we will concentrate on the 
$d \ln f/dt$ expression in Eq. (\ref{dlnf_MT}) from here onwards.
Combining Eqs. (\ref{dlnf_MT}),
 (\ref{dJorb_dt}) and (\ref{tau_mt}), a concise form of this expression can be
 written
%the Based on the above discussion and Eq.(\ref{dlnf_MT}), we can write
to indicate its behavior during
 both the inspiral \citep{Nel2001} and CMT \citep{Nel2004}
phases as:
%\footnote{The expression given in Eq.(\ref{fdot_eq}) above is identical to the 
%expressions given in \cite{Nel2001} for $g=0$and \cite{Nel2004} for , except in
% \cite{Nel2004} there is a typographical error in Eq.(12) where the term in the
% second parenthesis should be raised to the power of $-1$.}:
\begin{equation}
\frac{df}{dt} = \frac{3f}{8 \tau_\mathrm{chirp}}\biggl[2g - 1\biggr]
\label{fdot_eq}
\end{equation}
where,
\begin{eqnarray}
g &=& 0  \,  \qquad\qquad\qquad
(\textrm{inspiral phase}); \, \label{g-inspiral}  \\
g &=&  \frac{(1-q)}{\Delta\zeta}  \, \qquad \qquad (\textrm{CMT
phase}).\label{g-masstransfer}
\end{eqnarray}
Since $\Delta\zeta$ depends on both $M_\mathrm{tot}$ and $q$,
 there exists a critical $q = q_\mathrm{crit}(M_\mathrm{tot})$
for which $\Delta\zeta(M_\mathrm{tot}, q) = 0$.
% This value of
%$q_\mathrm{crit}$ represents the nature of mass transfer mechanism for DWDs.
For systems with
 $q > q_\mathrm{crit}(M_\mathrm{tot})$,  a phase of unstable mass transfer
 ensues and our
 present analysis becomes invalid in that regime. Hence
 $q_\mathrm{crit}(M_\mathrm{tot})$
represents the limiting mass ratio for a system to be in stable CMT phase.

\section{Detectability of DWD Systems}\label{SNR_section}

\subsection{Systems with Non-negligible Frequency Variations}

As we have discussed, the physical processes that drive the
evolution of DWD binaries operate on a ``chirp'' timescale, and
$\tau_\mathrm{chirp}$ is typically much longer than the operational
 time for LISA
(assumed one year here). Hence,
the time-variation of a given system's GW frequency $f(t)$ can be
well approximated by a truncated Taylor series expansion in time
and, using Eq.~(\ref{phaseDefinition}), the observed phase of the GW
signal $\phi_\mathrm{O}$ can be written in the form 
%{\bf [For Joel : I 
%am not referencing \cite{SVN05} here because this is not the expression given
%in their paper.]}
%\citep{SVN05},
%%\footnote{A similar discussion can be found in \cite{SVN05} where
%%higher order derivatives in the frequencies are considered to define
%%a parameter called breaking index which differentiates between the
%%physical mechanisms affecting the binary evolution.}
\begin{equation}
\phi_\mathrm{O}(t) = \phi_0 + 2\pi f_0 t + 2\pi\biggl[\frac{t^2}{2!} \dot f
+\frac{t^3}{3!} \ddot f + ...\biggr]
\, ,
\end{equation}
%\begin{equation}
%\phi_\mathrm{O}(t) = \phi_0 + 2\pi f_0 t + 2\pi
%\sum_{k=1}^{k_\mathrm{max}} \frac{t^{k+1}}{(k+1)!} f^{(k)} \, ,
%\end{equation}
where $f_0$ is the signal frequency at time $t=0$, 
$\dot f = df/dt$, $\ddot f = d^{2}f/dt^{2}$,
 and so on.
%the
%``spin-down parameters'' $f^{(k)} \equiv d^kf/dt^k (k = 1, \ldots,
%k_\mathrm{max})$.
If we truncate the Taylor series at $\dot f$
% If, for example, the Taylor series can be
%truncated at $k_\mathrm{max}=1$
 and this observed signal (O) is compared
to a computed template (C) that assumes a continuous-wave signal and
therefore has a phase that increases only linearly with time, 
$\phi_\mathrm{C}(t) = (\phi_0 + 2\pi f_0 t)$,
% then assuming
%that a sufficient degree of phase coherence is maintained if
% $\phi_\mathrm{O} - \phi_\mathrm{C} = \pi / 2$,
 the amount of time for
the O-C phase difference to reach $\pi/2$ 
%(assuming a sufficient degree of phase coherence is maintained until then)
 will be,
\begin{equation}
t_\mathrm{O-C} = (2 ~|\dot f|)^{-1/2} \, . \label{tO-Csimple}
\end{equation}
Substituting for $\dot f$ from Eq. (\ref{fdot_eq}), we obtain
%From Eqs.~(\ref{dlnf_inspiral_final}) and (\ref{dfdt_mt}) we see
%that, for both the inspiral and CMT phases of DWD evolutions, the
%first time-derivative of the frequency can be written in the form,
%\begin{equation}
%\dot f \approx \frac{3f_0}{8\tau_\mathrm{chirp}} \biggl[ 1 - 2g
%\biggr] \, , \label{genericFprime}
%\end{equation}
%where $g$ has been defined in connection with Eq.(\ref{dfdt_mt})
%\begin{eqnarray}
%g &=& 0  \,  \qquad\qquad\qquad
%(\textrm{inspiral phase}); \, \label{g-inspiral}  \\
%g &=&  \frac{(1-q_0)}{\Delta\zeta}  \, \qquad \qquad (\textrm{CMT
%phase}).\label{g-masstransfer}
%\end{eqnarray}
%Hence, we can write,
\begin{equation}
t_\mathrm{O-C} = \biggl(\frac{4\tau_\mathrm{chirp}}{3 |1-2g| f_0}
\biggr)^{1/2} = \biggl[ \frac{ 5}{48 \pi^2 |1-2g| } \biggl(
\frac{c}{rh_0 f_0^3} \biggr) \biggr]^{1/2} \,
.\label{tO-Cexpression}
\end{equation}

%As an illustration, in the top panel of Figure
%\ref{SNR_comparison} we have plotted the function
%$t_\mathrm{O-C}(f)$ for DWD binaries that lie along the $q=2/3$
%inspiral termination boundary  shown in Figure
%\ref{curved_parameter_space}. Over this entire range of frequencies,
%$t_\mathrm{O-C} \leq 1$ year; indeed, at the highest frequencies
%$t_\mathrm{O-C}$ drops well below one week. Combining this
%calculation of $t_\mathrm{O-C}$ with expression
%(\ref{EQ:reducedSNR}) produces the lower (red) curve in the
%bottom panel of Figure \ref{SNR_comparison}.  This curve
%provides a more realistic estimate of the SNR that DWD systems of
%this type (that lie at a distance of 10~kpc) will exhibit in LISA
%data if they are assumed to be continuous-wave sources. In the
%frequency range of $10^{-1}$ - $10^{-2}$ Hz, they will have roughly
%an order of magnitude lower SNR than one would estimate from a
%simple measurement of $\Delta\log h$ in Figure
%\ref{curved_parameter_space}.  For these systems, the higher SNR
%depicted by the upper (green) curve in Figure \ref{SNR_comparison}
%will be realized only if a proper inspiral template is used during
%data analysis to ensure that phase coherence of the signal is
%maintained over a full year of signal integration.

If the function $g$ in Eq.(\ref{tO-Cexpression}) is independent of
$h$ and $f$ --- as is the case for the inspiral phase of DWD
evolutions --- then curves of constant $t_\mathrm{O-C}$ in the
amplitude-frequency diagram will be straight lines having a slope of
$-3$. An example of this curve, assuming $t_\mathrm{O-C} = 1$ year,
is shown in Figure \ref{conclusion_figure} joining the 
two low frequency points.
% In Figure \ref{curved_parameter_space} we have drawn a line
%segment of slope $-3$ that identifies which inspiral systems have
%$t_\mathrm{O-C} = 1$ year.  Inspiral systems that lie below and to
%the left of this line segment have $t_\mathrm{O-C} > 1$ year, while
%systems that lie above and to the right have $t_\mathrm{O-C} < 1$
%year.
Any inspiral system that lies to the right of this constant
 $t_\mathrm{O-C}$ line
%regions identified in Figure \ref{curved_parameter_space}
 will lose
phase coherence in less than one year of observation if one assumes
that it emits continous-wave radiation.  An analogous one-year
demarcation boundary can be drawn for DWD binaries that are
undergoing a phase of stable, CMT by evaluating
Eq.~(\ref{tO-Cexpression}) using the function $g(q, M_\mathrm{tot})$.
%given by expression~(\ref{g-masstransfer}).
 Because this function
generally is of order unity,
% however (see Appendix
%\ref{appendix_A}),
 the one-year demarcation boundary for
mass-transferring systems is generally well-approximated by the line
segment that marks the one-year demarcation boundary for inspiral
systems.
% On the upside, the $t_\mathrm{O-C}$ curve indicates that
LISA will be able to measure frequency evolution for DWD systems that 
lie to the right of the $t_\mathrm{O-C} = 1$ year line and, as will be discussed in the 
following section, it will then be possible to determine the distances
and masses for these systems. 
%within one year of observation time.
 The downside is that
millions of DWD systems will lie to the left of the one year 
$t_\mathrm{O-C}$ curve 
in the low frequency region, for which frequency evolution cannot be measured.
\subsection{Determination of Distances and Masses} \label{q_and_mtot}

An analysis of a one-year-long LISA data stream that utilizes a
proper set of frequency-varying strain templates should be able to
determine the rate at which the GW frequency 
%and, hence, the orbital
%frequency
 is changing in both inspiral and mass-transfering DWD systems.
% that are identified as sources
%inside the triangular-shaped regions of Figure
%\ref{curved_parameter_space}.
 When used in conjunction with the
measurement of $h_\mathrm{norm}$ and $f$, an accurate measurement of
$\dot f$ for any source should permit a determination of the
distance to the source $r$ and should give information about the
chirp mass $\mathcal{M}$ or the individual component masses of the
binary system, as follows.

Equation (\ref{h_f_relationship}) provides a relation between the
three unknown binary system parameters $r, M_\mathrm{tot}$ and $q$,
and the experimentally measurable parameters $f$ and
$h_\mathrm{norm}$, namely,
\begin{eqnarray}
\frac{M_\mathrm{tot}^5}{r^3} \biggl[ \frac{q}{(1+q)^2}\biggr]^3 &=&
\frac{\mathcal{M}^5}{r^3} =  \frac{c^{12}}{2^6 \pi^2 G^5} \biggl[
\frac{h_\mathrm{norm}^3}{ f^2} \biggr] \, .\label{Unknowns1}
\end{eqnarray}
A second relation between the unknown astrophysical parameters and
measurable ones is provided by combining the expression for
$\dot f$ in Eq.~(\ref{fdot_eq}) with the definition of
$\tau_\mathrm{chirp}$ given in Eq.~(\ref{tau_chirp}). Specifically,
we obtain,
\begin{equation}
r (1-2g)  = \frac{5c}{24 \pi^2} \biggl[
\frac{\dot f}{h_\mathrm{norm}f^3}\biggr] \, , \label{Unknowns2}
\end{equation}
%where $g$ is defined in Eq's. (\ref{g-inspiral}) and (\ref{g-masstransfer}).
With only two equations, of course, it is not possible to
uniquely determine all three of the binary's primary system
parameters. However, in the inspiral phase $g = 0$, so $\mathcal{M}$ and $r$
 can be determined as was shown in \cite{Schutz86}.

% During the inspiral phase of a DWD evolution, however,
%$g=0$, so a fortunate situation arises.  Equation (\ref{Unknowns2})
%drops its explicit dependence on the system mass to give a clean
%determination of $r$.  But once $r$ has been determined,
%Eq.~(\ref{Unknowns1}) gives only the chirp mass $\mathcal{M}$,
%rather than giving $M_\mathrm{tot}$ and $q$ separately. This is a
%familiar result \citep{Schutz86}.

During the CMT phase of an evolution, the function
$g(M_\mathrm{tot},q)$ is nonzero so Eq.~(\ref{Unknowns2}) does not
provide an explicit determination of $r$.  However, the requirement
that $R_d = R_L$ provides an important additional physical
relationship between the unknown astrophysical parameters and
measurable ones. Specifically, by setting $R_d$ from
Eq.~(\ref{mass_radius}) equal to $R_L$ from Eq.~(\ref{roche_radius})
and using Kepler's law to write $a$ in terms of $f$, we obtain,
\begin{eqnarray}
\biggl[\frac{R_\mathrm{\sun}^3}{ GM_\mathrm{\odot}} \biggr]^{1/2} f
&=& \biggl[ \pi^2 (0.0114)^3 \frac{M_\mathrm{ch}}{M_\odot}
\biggr]^{-1/2} \frac{M_\mathrm{tot}}{M_\mathrm{\odot}} \biggl(
\frac{q}{1+q} \biggr)H(M_d,q)    \, , \label{Unknowns3}
\end{eqnarray}
where,
\begin{eqnarray}
H(M_d,q) &\equiv&    \biggl( \frac{1+q}{q} \biggr)^{1/2}\biggl[
\frac{0.49 \ q^{2/3}}{0.6 \ q^{2/3} + \ln(1 + q^{1/3})}\biggr]^{3/2}
\biggl[1 - \biggl(\frac{M_d}{M_\mathrm{ch}}
\biggr)^{4/3}\biggr]^{-3/4} \nonumber \\
&& \times \biggl[1+3.5\biggl(\frac{M_d}{M_\mathrm{p}}\biggr)^{-2/3}+
\biggl(\frac{M_d}{M_\mathrm{p}}\biggr)^{-1}\biggr]  \,
.\label{define_H}
\end{eqnarray}
Hence, taken together, Eqs.~(\ref{Unknowns1})-(\ref{Unknowns3}) can
be used to determine all three primary system parameters -- $r$,
$M_\mathrm{tot}$, and $q$ -- from the three measured quantities,
$h_\mathrm{norm}, f$, and $\dot f$. 
% (We stress that this method of
%determining the values of the primary system parameters is only
%valid in situations where $q < q_\mathrm{crit}(M_\mathrm{tot})$, as
%explained in Appendix \ref{appendix_A}.)
\clearpage
\begin{figure}[!hbp|t]
\centering
\epsscale{1}
\plotone{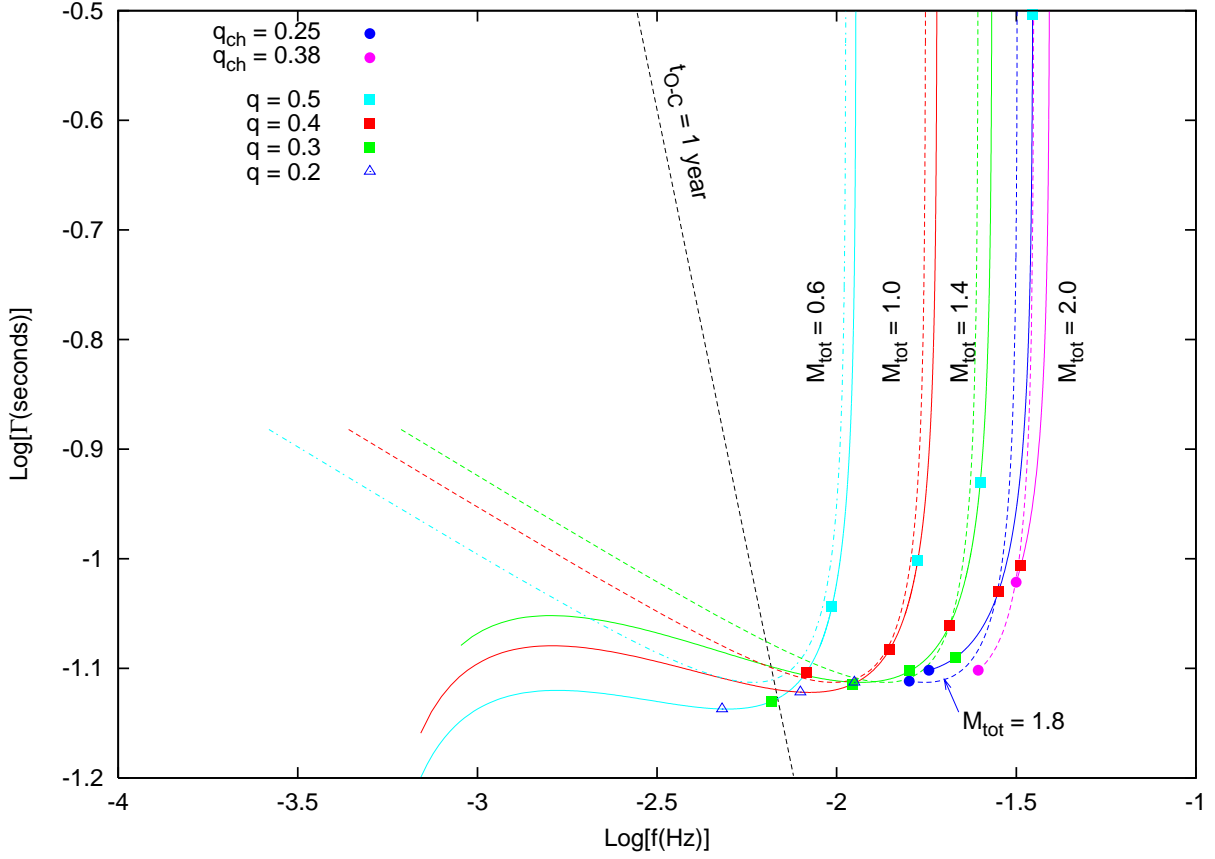}
\hfill \caption{Plot of $\log \Gamma$ versus $\log f$, where $\Gamma
\equiv \{[-\dot f]^3/f^{16}\}^{1/10}$, for stable CMT trajectories
having five different system masses ranging from $0.6 M_\odot$ to
$2.0 M_\odot$, as labeled. Two trajectories are shown for each value
of $M_\mathrm{tot}$: Dashed curve is the analytic solution provided
by Eqs.~(\ref{Pac_q_manuscript}) and (\ref{Pac_M_manuscript}); solid
curve shows the numerical solution obtained from
Eqs.~(\ref{Unknowns1}) through (\ref{Unknowns3}). Different values
of $q$ are identified by various symbols along each trajectory; note
that the trajectories with $M_\mathrm{tot} = 2 M_\odot$ and $1.8
M_\odot$ terminate at values of $q = q_\mathrm{ch} = 0.38$ and
$0.25$, respectively. } \label{Fig:ffdot_Pac}
\end{figure}
\clearpage
We are unable to solve this set of equations analytically due to the
complexity of the functions $g(M_\mathrm{tot},q)$ and $H(M_d,q)$.
However, the formulae that \cite{Paczynski67} adopted for $R_d(M_d)$
and $R_L(q)$ 
%(see Appendix \ref{Appendix_B})
 lead to much simpler
expressions for both of these functions, namely, $g =
[\frac{3}{2}(1-q)/(2-3q)]$ and $H=1$. 
%As is shown in Appendix
%\ref{Appendix_B},
 In this case
Eqs.~(\ref{Unknowns1})-(\ref{Unknowns3}) can be combined to give
%Eq.~(\ref{Pac_q_manuscript}), which provides the following analytic expression
%for the mass ratio
 $q$ in terms of $f$ and $\dot f$:
\begin{equation}
q^2 (1+q)\biggl(1-\frac{3}{2}q\biggr)^3 = \biggl[\frac{2^{12}3^3
\pi^8 \alpha^5}{5^3 c^{15}} \biggr]\frac{f^{16}}{[-\dot f]^3} \, ,
\label{Pac_q_manuscript}
\end{equation}
where $\alpha \equiv 0.0141 (GM_\odot R_\odot^3)^{1/2}$.  Once $q$
is known, $r$ can be obtained using Eq.(\ref{Unknowns2}) in
conjunction with Paczy\'nski's $g(q)$ relation; then
$M_\mathrm{tot}$ can be obtained from Eq.(\ref{Unknowns1}).
%Specifically, from relations (\ref{Pac_rh}) and (\ref{Pac_Mtot}) we
%obtain, respectively,
\begin{eqnarray}
r &=& \frac{5c}{24\pi^2} \biggl[\frac{-\dot f}{h_\mathrm{norm}f^3}
\biggr]
(2-3q) \, ; \label{Pac_r_manuscript} \\
M_\mathrm{tot} &=& \biggl[ \frac{5^3 c^{15}}{2^{15}\cdot 3^3 \pi^8
G^5} \biggr]^{1/5} \biggl\{ \frac{(1+q)^{6}(2-3q)^3}{q^{3}} \cdot
\frac{[-\dot f]^{3}}{f^{11}} \biggr\}^{1/5} \, .
\label{Pac_M_manuscript}
\end{eqnarray}
For any $M_\mathrm{tot} \leq  2 M_\mathrm{ch}$, these three
equations are valid for mass ratios over the range $0 < q < 2/3$
because, for Paczy\'nski's model, $q_\mathrm{crit} = 2/3$
independent of $M_\mathrm{tot}$.
%(see Appendix \ref{appendix_A}).

The solid curves in Figure \ref{Fig:ffdot_Pac} illustrate results
obtained numerically from a self-consistent solution of
Eqs.~(\ref{Unknowns1})-(\ref{Unknowns3}); the dashed curves
illustrate results obtained analytically from expressions
(\ref{Pac_q_manuscript}) and (\ref{Pac_M_manuscript}). Across the
parameter domain defined by the two observables $\log(f)$ and
$\log(\Gamma)$ --- where
\begin{eqnarray}
\Gamma &\equiv& \{[-\dot f]^3/f^{16}\}^{1/10}   \, ,
\label{define_gamma}
\end{eqnarray}
is measured in seconds --- each curve traces a constant
$M_\mathrm{tot}$ ``trajectory'' with  $q$
varying along each curve, as indicated. At high frequencies, each
curve begins at a value of $q$ that is slightly below $q_\mathrm{crit}$;
%$q_\mathrm{crit}$;
 at low frequencies, the curves have been extended
down to $q=0.05$, unless $M_\mathrm{tot} > M_\mathrm{ch}$, in which
case the curve has been terminated at the value $q = q_\mathrm{ch}$,
as given by Eq.~(\ref{qChandra}). The general behavior of these
curves can best be understood by analyzing analytic expression
(\ref{Pac_q_manuscript}).  Over the relevant range of mass ratios $0
\leq q \leq  2/3$, the analytic function,
\begin{eqnarray}
\Gamma_\mathrm{anal} &=& 0.0521 \biggl[ q^2
(1+q)\biggl(1-\frac{3}{2}q\biggr)^3 \biggr]^{-1/10}
~~\mathrm{seconds} \, ,
\end{eqnarray}
reaches a minimum value ($\Gamma_\mathrm{min} =
0.077~\mathrm{seconds}$) when $q = q_\mathrm{extreme}$, where
\begin{equation}
q_\mathrm{extreme} \equiv \frac{1}{12}(\sqrt{41} -3) = 0.2836 \, .
\label{q_extreme}
\end{equation}
Moving from high frequency to low frequency along each
$M_\mathrm{tot}$ ``trajectory,'' the function $\Gamma$ steadily
drops until $q = q_\mathrm{extreme}$ and $\Gamma =
\Gamma_\mathrm{min}$. (This behavior holds for the solid curves as
well as the dashed curves, although the precise values of
$\Gamma_\mathrm{min}$ and $q_\mathrm{extreme}$ are different for
each solid curve.)  When $q$ drops below $q_\mathrm{extreme}$ [based
on the function $q_\mathrm{ch}$, this will only happen along curves
for which $M_\mathrm{tot} < (1+q_\mathrm{extreme})M_\mathrm{ch} =
1.85 M_\odot$], each curve climbs back above $\Gamma_\mathrm{min}$,
reflecting the fact that Eq.~(\ref{Pac_q_manuscript}) admits two
solutions over the relevant range of mass ratios. This, in turn,
implies that for mass-transferring DWD systems that have $\log(f)<
-1.74$, a measurement of $\dot f$ will generate two possible
solutions -- rather than a unique solution -- for the pair of key
physical parameters $(M_\mathrm{tot},q)$.

Once LISA has measured $\dot f$ as well as $f$ for a given DWD
system, Figure \ref{Fig:ffdot_Pac} provides a graphical means of
determining the values of $M_\mathrm{tot}$ and $q$ for the system,
assuming it is undergoing a phase of stable CMT.  We do not expect
that LISA will probe the entire parameter space depicted in this
figure, however.  As discussed above, we expect that LISA will only
be able to detect frequency changes in systems for which
$t_\mathrm{O-C} \lesssim 1~\mathrm{yr}$.  Using expression
(\ref{tO-Csimple}), this means that LISA will only be able to
measure $\dot f$ for systems that have,
\begin{equation}
\Gamma \gtrsim 2.57 \times 10^{-5} f^{-8/5}~\mathrm{seconds} \, .
\end{equation}
The dashed black line in Figure \ref{Fig:ffdot_Pac} with a slope of
$-8/5$ that is labeled ``$t_\mathrm{O-C} = 1~\mathrm{year}$'' shows
this boundary; the parameter regime that can be effectively probed
by LISA lies above and to the right of this line.

\section{Conclusions}
\subsection{Principal Findings}
\begin{figure}[!hbp|t]
\centering
\plotone{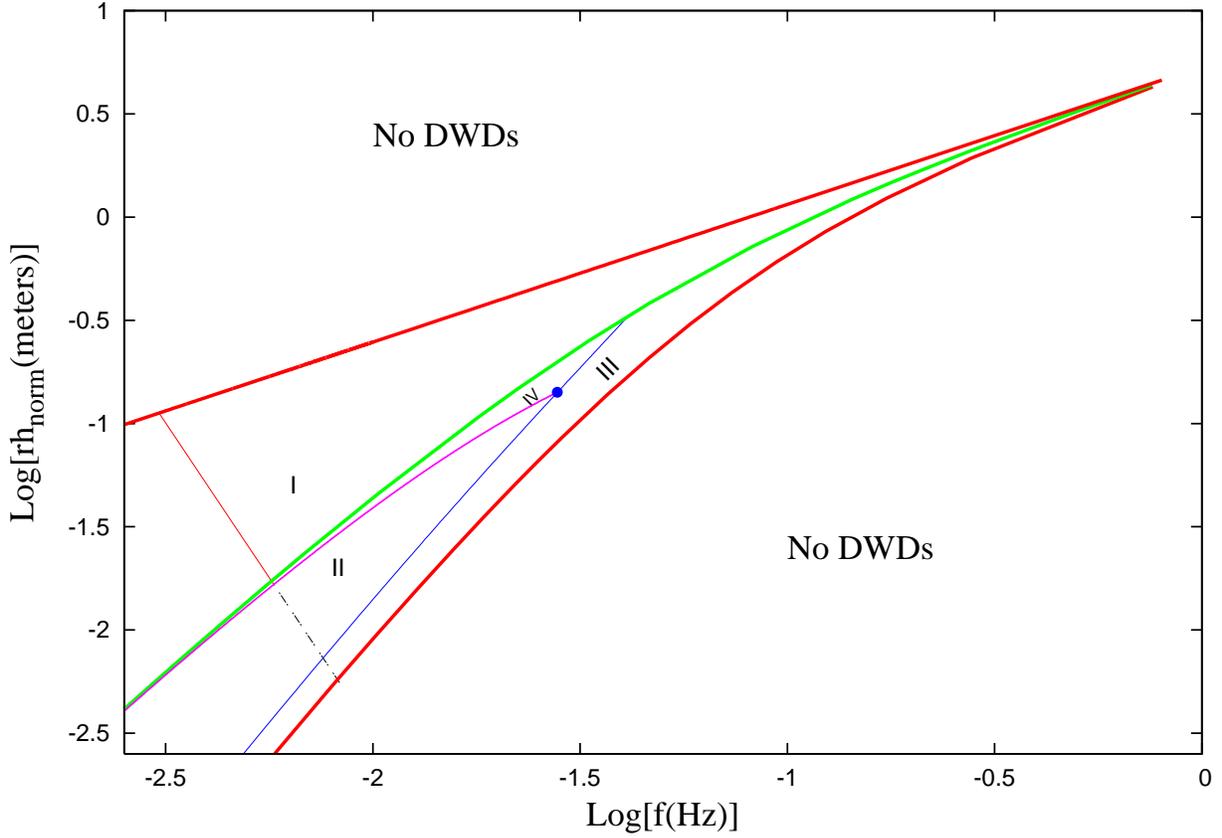}
\hfill \caption{DWD population boundaries are drawn in LISA's
``absolute'' amplitude-frequency diagram; five boundary curves
(identified as curves {\bf A}-{\bf E} in the text) divide the
diagram into four principal population sub-domains, numbered
$I$-$IV$. Only inspiraling systems will be found in {\it Region I};
systems in a phase of stable CMT will only be found in {\it Region
II}; {\it Region III} contains inspiraling systems that will
eventually encounter a phase of unstable mass transfer; and systems
in {\it Region IV} are progenitors of Type Ia supernovae. {\it Zones
of Avoidance}: As indicated, no DWD systems will be found in the
regions that lie outside of the two thick red curves (identified in
the text as boundary curves {\bf A} and {\bf B}). The straight line
of slope $-3$ drawn in the lower left-hand quadrant (identified as
curve {\bf T} in the text) is the $t_\mathrm{O-C} = 1~\mathrm{yr}$
demarcation line; LISA will only be able to measure $\dot f$ values
for systems that lie above and to the right of this line.}
%\hfill \caption{DWD Populations and SN progenitors. The two thick
%bounding curves are for $q=1$ and all the DWD systems that are
%interesting for LISA are confined within this region.  Systems which
%lie above the line joining the two low frequency points of these
%bounding curves have an appreciable frequency change such that they
%lose phase coherence in less than one year if one assumes that they
%emit continous wave radiation. Region I has only inspiral sources,
%region II has both inspiral and mass transferring systems and most
%of the known AM CVn systems at present will lie in this region (with
%much lower amplitude and frequency),
% region III systems, if they are mass tranferring binaries,
% undergo unstable mass transfer period and may have a violent ending and
%region IV mass transferring systems are progenitors of Type Ia
%supernovae. }
 {\label{conclusion_figure}}
\end{figure}

Once the distance has been determined to individual LISA sources, as
was outlined in \S\ref{q_and_mtot}, it will be possible to place them
 in an ``absolute''
amplitude-frequency diagram, that is, in a
$\log(rh_\mathrm{norm})-\log(f)$ diagram. The location of individual
sources in
%, and the distribution of the entire collection of LISA
%sources across,
 such a diagram should help us understand a great
deal about our Galaxy's DWD population.  Our expectation is that
systems in different evolutionary states will fall into several
distinct sub-domains of LISA's ``absolute'' amplitude-frequency
diagram, and the diagram will exhibit natural zones of avoidance as
well.  Our models of inspiral and CMT systems permit us to predict
%in a quantitative manner
 where the boundaries will lie between these
various population sub-domains.  As depicted in Figure
\ref{conclusion_figure}, the principal sub-domains can be identified
as follows.

\clearpage
\clearpage

\begin{itemize}
\item {\it Zones of Avoidance}:  No DWD systems will be found
in the region above the boundary line that is defined by an inspiral
trajectory with $K=1$; according to Eq.~(\ref{h_f_relationship}),
this boundary is defined by the expression,
\begin{equation}
\log (rh_\mathrm{norm}) = 0.731 + \frac{2}{3} \log f \, . \qquad
{\bf Boundary ~A} \label{Boundary_A}
\end{equation}
Also, no DWD systems will be found in the region below the bounding
curve that is defined by the inspiral termination boundary for
systems with $q=1$; over the region of parameter space shown in
Figure \ref{conclusion_figure} this bounding curve is given
approximately by the expression,
\begin{eqnarray}
\log (rh_\mathrm{norm}) \approx  0.703 &+& 0.637~ \log f
-0.017~ (\log f)^2 \nonumber \\
&+& 0.298~ (\log f)^3 +0.061~ (\log f)^4 \, . \qquad {\bf Boundary
~B} \label{Boundary_B}
\end{eqnarray}
Both of these population boundaries have been drawn as solid red
curves in Figure \ref{conclusion_figure}.
\end{itemize}
Detached, inspiraling DWD systems will be found throughout the
domain whose upper and lower borders are defined, respectively, by
boundary curves {\bf A} and {\bf B}.  This region of parameter space
can be subdivided into the following two principal population
sub-domains.
\begin{itemize}
\item {\it Region I}:  Only inspiraling DWD systems will
be found in the region of parameter space that is bounded above by
curve {\bf A} and below by the locus of points that identifies
semi-detached systems for which $M_a = M_\mathrm{ch}$ (the green
solid curve in Figure \ref{conclusion_figure}); over the region of
parameter space shown in Figure \ref{conclusion_figure}, this curve
is given approximately by the expression,
\begin{eqnarray}
\log (rh_\mathrm{norm}) \approx 0.761 &+& 1.005~\log f + 0.700
(\log f)^2 + 0.700 (\log f)^3 \nonumber \\
&+& 0.214 (\log f)^4 + 0.023 (\log f)^5 \, . \qquad {\bf Boundary
~C} \label{Boundary_C}
\end{eqnarray}

\item {\it Region II}:
Mass-transferring DWD systems will only be found in the region of
parameter space that is bounded above by curve {\bf C} and below by
curve {\bf B}.
\end{itemize}
Region II can be further subdivided into two significant population
sub-domains as follows:
\begin{itemize}
\item {\it Region III}:
An inspiraling DWD system that has $q > q_\mathrm{crit}(M_\mathrm{tot})$ value
%> q_\mathrm{crit}(M_\mathrm{tot})
 will encounter an unstable --
rather than a stable -- phase of mass-transfer when the less massive
star initially makes contact with its Roche lobe.  The region of
parameter space in which these systems will be found at the onset of
mass-transfer is bounded below by curve {\bf B} and above by two
intersecting curve segments:  At $\log f \gtrsim -1.38$, curve {\bf
C} defines the upper boundary; for $\log f \lesssim -1.38$, the
relevant upper boundary has been drawn as a blue solid curve in
Figure \ref{conclusion_figure} and is given approximately by the
expression,
\begin{eqnarray}
\log (rh_\mathrm{norm})\approx 2.141 &+& 1.686 (\log f)  \nonumber \\
&-& 0.141 (\log f)^2 + 0.007 (\log f)^3 \, . \qquad {\bf Boundary
~D} \label{Boundary_D}
\end{eqnarray}

\item {\it Region IV}:
DWD systems that enter a phase of stable CMT with a total system
mass $M_\mathrm{tot} > M_\mathrm{ch}$ will be found in a region of
parameter space that is bounded above by curve {\bf C} and below by
two intersecting curve segments:  For $-1.55 \leq \log f \leq
-1.38$, curve {\bf D} defines the lower boundary; for $\log f \leq
-1.55$, the relevant lower boundary has been drawn as a pink solid
curve in Figure \ref{conclusion_figure} and is given approximately
by the expression,
\begin{eqnarray}
\log (rh_\mathrm{norm})\approx &-& 1.381 - 2.108 (\log f) \nonumber
\\
&-& 1.394 (\log f)^2 - 0.167 (\log f)^3 \, . \qquad {\bf Boundary
~E}
\end{eqnarray}
Boundary curve {\bf E} is defined by the stable CMT trajectory for a
system with $M_\mathrm{tot} = M_\mathrm{ch}$.
\end{itemize}

An additional demarcation line has been drawn in Figure
\ref{conclusion_figure} that is associated with the $t_\mathrm{O-C}
= 1$ yr boundary.
%shown originally in Figure
%\ref{curved_parameter_space}.
 This ``time boundary'' line {\bf T}
is defined by setting $g=0$ and $t_\mathrm{O-C} = 1$ yr in
Eq.~(\ref{tO-Cexpression}), that is, it is given by the expression,
\begin{eqnarray}
\log (rh_\mathrm{norm})= -8.498 - 3 \log f \, . \qquad {\bf Boundary
~T}
\end{eqnarray}
LISA will be unable to determine distances to DWD systems that lie
below and to the left of this demarcation line because their orbital
periods and associated GW frequencies are changing so slowly that
LISA will be unable to measure with confidence the value of
$\dot f$ for these systems.  Studies of DWD
populations will therefore benefit most from the data that LISA
collects on systems that lie above and to the right of boundary line
{\bf T}. The segment of this line that bounds {\it Region II} has been
drawn as a dashed line to emphasize that it is only an approximate one year
boundary for mass-transferring systems.

%The chirp mass $\mathcal{M}$ can be determined for inspiraling DWD
%systems once LISA has measured $f$ and $\dot f$ \citep{Schutz86};
%% specifically, a
%%combination of Eqs.~(\ref{Unknowns1}) and (\ref{r_final_inspiral})
%%gives,
%%\begin{eqnarray}
%%\mathcal{M} &=& \biggl[ \frac{5^3 c^{15}}{2^{15}\cdot 3^3 \pi^8 G^5}
%%\biggr]^{1/5} \biggl\{ \frac{[f^{(1)}]^{3}}{f^{11}} \biggr\}^{1/5}
%%\, .
%%\end{eqnarray}
%For DWD systems that are in a stable CMT phase of evolution,
%% our model predicts that
% the degeneracy of the chirp mass can be broken.
%In principle,  measurements of $f$ and $\dot f$ permit
%a determination of both $M_\mathrm{tot}$ and $q$ and, hence, a
%determination of the masses of the individual stars in the binary
%system.  For CMT systems, Eqs.~(\ref{Pac_M_manuscript}) and
%(\ref{Pac_q_manuscript}) provide approximate expressions for
%$M_\mathrm{tot}$ and $q$, respectively, and Figure
%\ref{Fig:ffdot_Pac} provides a graphical means of determining these
%two key system parameters.

\subsection{Discussion}

The particular boundaries {\bf A}-{\bf E} of the population domains
that are depicted in Figure \ref{conclusion_figure} arise as a
consequence of the specific mass-radius relationship
(Eq.~\ref{mass_radius}) and, hence, the equation of state that we
have chosen to use to describe the structure of individual white
dwarfs. Because it is generally believed that the 
mass-radius relationship given in Eq.(\ref{mass_radius})
represents the properties of white dwarfs
quite well, we expect that the population sub-domains and zones of
avoidance shown in Figure \ref{conclusion_figure} will map well onto
LISA's observationally determined ``absolute'' amplitude-frequency
diagram. Systems that are found on the ``wrong'' side of a given
boundary -- for example, any system that lies in one of the zones of
avoidance, or CMT systems that are found in Region I -- will be of
particular interest because they may provide evidence that the
equation of state that we adopted
%utilized by \cite{N72}
 is not general enough to
properly describe this stellar population
 Two examples suffice to
illustrate this point.  First, in the presence of tidal stresses,
the donor stars in mass-transferring DWD systems may be hotter than
assumed here \citep{Bild2002, Deloye2003}, which will
affect the frequency at which the donor fills the Roche lobe.  Consequently,
the inspiral termination boundary {\bf B} in Figure 3 will change. Second,
 because individual neutron stars obey
an entirely different mass-radius relationship and generally seem to
have masses $\gtrsim M_\mathrm{ch}$, inspiraling double neutron-star
systems will likely be distinguishable from DWD systems because they
will lie in the zone of avoidance above boundary curve {\bf A}, as
defined above.

DWD systems that are undergoing a phase of stable, CMT and that are
found to reside in Region IV of the ``absolute'' amplitude-frequency
diagram can be identified as progenitors of Type Ia supernovae.
Efforts to better understand the origin of supernova explosions in
old stellar populations will especially benefit from follow-up
studies that identify the optical (or UV or x-ray) counterparts to
these Region IV systems.  Inspiraling systems that are found to
reside in Region III may prove to be equally interesting candidates
for follow-up studies.  Detached DWD systems in Region III are
destined to enter a phase of unstable mass transfer (likely
accompanied by super-Eddington accretion, see \cite{GPF06}) that
will significantly transform the system's properties on a dynamical,
rather than a chirp, timescale. Such rapid mass-transfer events may
lead to merger of the binary components, perhaps followed by an
explosion.

As stated above, we are confident that the population sub-domains
and zones of avoidance shown in Figure \ref{conclusion_figure} will
map well onto LISA's observationally determined ``absolute''
amplitude-frequency diagram.
% because their boundaries are fixed
%largely by the time-independent properties of the \cite{N72}
%mass-radius relationship.
 We are less confident about the degree to
which LISA's measurements of $\dot f$ will match the values that
are predicted by our simplified model of the slow, orbit-averaged
time-evolution of DWD systems (as displayed, for example, in Figure
\ref{Fig:ffdot_Pac}). Mass-transferring binary systems, in
particular, are notoriously messy laboratories.  For example,
significant and unexplained variations in the mass-transfer rate can
arise in an individual system over times that are much shorter than
the GR-driven evolutionary time scale; this can introduce
significant short-term variations in $\dot f$. Also, magnetic
fields can be effective at carrying away mass and angular momentum
from a system, thereby violating our assumption of conservative mass
transfer. V407 Vul and RX J0806+1527 \citep{MN2005} provide perhaps the
best examples of the type of confusion that is likely to arise when
attempts are made to extract measurements of $\dot f$ from a
one-year-long LISA data stream. These are optically identified,
mass-transferring binaries that are thought to be DWD systems
because their orbital periods are less than ten minutes. The best
available measurements of period variation in both of these systems
indicate that their orbits are slowly shrinking, rather than slowing
growing larger as would be predicted by our model of stable CMT.
Although, in the mean, the long-term evolutionary behavior of these
systems is likely to agree with the predictions of our model,
fluctuations about the mean that occur on a timescale that is short
compared to $\tau_\mathrm{chirp}$ may totally confound our ability
to interpret LISA's measurements of $\dot f$. On the bright side,
if a significant number of LISA sources exhibit noticeable
deviations away from the mean behavior predicted by our simplified
model, in the end we are likely to gain a more complete
understanding of the evolution of such systems.

%There may be a large number of individual DWD systems for which LISA
%is able to confidently measure $h_\mathrm{norm}$ and $f$, but not
%$\dot f$. Without a direct measure of $\dot f$, it will not be
%possible to determine distance to these systems.
%Nevertheless, the population boundaries that we have identified may
%provide useful limits on $r$ for such systems. Consider, for
%example, a system whose measured GW frequency is $0.1$ Hz.  Because
%the system must lie below curve {\bf A} and above curve {\bf B} in
%the ``absolute'' amplitude-frequency diagram,
%Eqs.~(\ref{Boundary_A}) and (\ref{Boundary_B}) tell us that $r$ must
%fall in the range,
%\begin{eqnarray}
%0.65 h_\mathrm{norm}^{-1} \leq r (\mathrm{meters}) \leq 1.16
%h_\mathrm{norm}^{-1} \, .
%\end{eqnarray}
%If there is independent evidence that such a system is in the CMT
%phase of its evolution ({\it e.g.}, the system is known from optical
%observations to be accreting) then a tighter constraint on the
%distance can be obtained by using curve {\bf C} rather than curve
%{\bf A} as the upper population boundary.  Because curves {\bf A},
%{\bf B} and {\bf C} diverge from one another at lower frequencies,
%the uncertainty in such estimates of $r$ will necessarily be larger
%for DWD systems with lower GW frequencies.

\acknowledgements

%We appreciate the advice of an anonymous referee, who urged us to shorten the
%original manuscript making our key point more succinct.
We acknowledge numerous very useful discussions that we have had
with J. Frank, V. Gokhale, P. Motl, and X. Peng about the
evolutionary behavior of DWD systems.  We appreciate the
encouragement that we received from S. Larson and M. Benacquista to
push this work to completion after they reviewed a preliminary
presentation of our population boundary diagram.  We also appreciate the advice
 of an anonymous referee, who urged us to shorten the
original manuscript making our key point more succinctly. This work has
been supported, in part, by NASA grant NAG5-13430 and by NSF grants
AST-0407070 and PHY-0326311.

\end{document}